\documentclass[twocolumn,showpacs,preprintnumbers,amsmath,amssymb,floatfix]{revtex4}

\usepackage{epsfig,psfrag}
\usepackage{dcolumn}
\usepackage{bm}
\usepackage{graphicx}
\usepackage{color}

\begin{document}

\title{Interaction-induced harmonic frequency mixing in quantum dots}
\author{M.~Thorwart,$^1$ R. Egger,$^1$ and A.O. Gogolin$^{1,2}$}
\affiliation{$^1$~Institut f\"ur 
Theoretische Physik, Heinrich-Heine-Universit\"at,
D-40225  D\"usseldorf, Germany \\
$^2$~Department of Mathematics, Imperial College, 180 Queen's Gate,
London SW7 2BZ, UK}
\date{\today}

\begin{abstract}
We show that  harmonic frequency mixing in quantum dots coupled to two leads 
under the influence of time-dependent voltages of different frequency
is dominated by interaction effects.  This offers a unique and
direct spectroscopic tool to access correlations,
and holds promise for efficient frequency mixing in nano-devices. 
Explicit results are provided for an Anderson dot and for a 
molecular level with phonon-mediated interactions.  
\end{abstract}

\maketitle

The nonlinear mixing of two signals with different 
frequencies is a widespread and important concept used in many areas
of physics.  Frequency mixers have been extensively used for a long time in 
microelectronics \cite{pozar}, where a diode provides the required 
nonlinearity,  and in  nonlinear optics 
(three- and four-wave mixing) \cite{mukamel}. 
Submillimeter-wave heterodyne signal detection  
at the quantum noise limit \cite{Tucker85},
based on the nonlinear current-voltage
characteristics of a superconducting tunnel junction (STJ),
is ubiquitous and has generated an enormous boost in radioastronomy. 
Apart from STJs, however, frequency mixing in quantum-coherent 
mesoscopic or nano-devices has been experimentally realized only for
a three-terminal ballistic junction \cite{Xu07} and for the 
single-electron transistor \cite{Cleland,Buehler},
the latter possessing a broad and tunable range of frequencies and bandwidths.  

In this work, we give the quantum theory of frequency mixing
in quantum dots -- serving as prototype examples for 
nano-devices -- in the presence of electron-electron (e-e) 
or electron-phonon (e-ph) interactions. 
Mesoscopic systems driven by ac fields are
of major interest in condensed matter physics 
\cite{platero,hanggi,jauho,bruder,buttiker,ackondo,schiller,goldin1,wang,fransson,polianski,polianski2},
but a quantum theory of mixing has only been 
formulated for STJs \cite{Tucker85}.
We consider a multi-tone setup, where the dot is attached to 
source and drain contacts (``leads'') with time-dependent (ac) voltages
of different drive frequencies $\omega_{L/R}$ \cite{footf}, and 
an additional gate is capacitively coupled to the dot. 
A quantity of main interest is the time-dependent 
current $I(t)$ through the device, expanded in a Fourier series as 
\begin{equation}\label{gen}
I(t) = {\rm Re} \sum_{n,m=-\infty}^\infty e^{-i\omega_{nm}t} I_{nm}, 
\end{equation}
with frequencies $\omega_{nm}=n\omega_L+m\omega_R$. 
By determining  the complex-valued $I_{nm}$, 
we derive general conditions under which $I(t)$ exhibits 
harmonic frequency mixing.  A mixing current can arise only if there exists 
at least one pair $(n>0,m\ne 0)$ 
with {\sl mixing amplitude} $J_{nm}\equiv I_{nm}+I^*_{-n,-m}\ne 0$ 
(the star denotes complex conjugation).  
In order to establish the importance of interaction
physics on frequency mixing, we explicitly compute 
the $J_{nm}$ for two important models, namely (i)
for on-site e-e interaction $U$
in a spin-degenerate single-level dot (Anderson impurity model), and 
(ii) for a spinless level with e-ph coupling
to an Einstein phonon, describing transport through 
vibrating molecules \cite{flensberg,review}.

Before turning to derivations, let us briefly summarize our
main findings in simple qualitative terms.
Taking the standard wide-band limit (WBL) for the leads, 
we find that no frequency mixing is possible in the absence of interactions.  
This is a striking and unexpected result that we shall discuss in some detail.
Once interactions are present, however, one  will
generally encounter mixing, $J_{nm}\ne 0$ \cite{footi}.
Therefore {\sl harmonic frequency mixing is dominated by interactions,} 
and thus provides a highly sensitive novel spectroscopic tool of 
correlations.  Importantly, we show that even in the linear response regime 
(dc bias voltage $V\to 0$) frequency mixing occurs in general, 
where the nonlinearity required for mixing
is now generated by interactions.  This implies that one may be able
to operate such a quantum-coherent nano-device as frequency mixer 
with reduced dissipation and shot noise.

We here take the left/right ($\alpha=L/R=\pm$) contact's single-particle
energies as time-dependent, $\epsilon_{k,\alpha}\to \epsilon_{k,\alpha}
\pm eV/2 + V^{ac}_{\alpha} \cos(\omega_{\alpha} t)$,
with dimensionless drive amplitudes $a_\alpha= V^{ac}_\alpha/\omega_\alpha$
(we put $\hbar=k_B=1$).
The time-dependent current $I(t)=[I_L(t)-I_R(t)]/2$
is then expressed in terms of 
$\Delta_\alpha(t) = a_\alpha \sin(\omega_\alpha t)$ 
and Fermi functions $f_\alpha(\omega)=f(\omega-\alpha eV/2)$ 
describing the respective contact,  
and the exact but unknown retarded (lesser) Green's functions (GF)
${\bf G}^r(t,t')$  (${\bf G}^<(t,t')$) of the dot \cite{jauho}.
A time-dependent gate voltage then produces a similar term $\Delta_g(t)$.
For simplicity, we assume that the tunneling amplitudes connecting the 
dot to the leads are not modulated by time-dependent
voltages, and correspond to energy-independent
hybridization matrices ${\bf \Gamma}_\alpha$
in the space of spin and dot level indices.
By a gauge-invariant generalization of Ref.~\cite{jauho}, 
the time-dependent currents $I_\alpha(t)$ for an interacting multi-level
quantum dot can be written in the form
\begin{eqnarray} \label{general}
I_\alpha(t) &=& - 2e {\rm Im Tr} [{\bf \Gamma}_\alpha 
{\bf G}^<(t,t)] \\ \nonumber
&-& 2e {\rm Im Tr}  \int dt' \int \frac{d\omega}{2\pi} f_\alpha(\omega)
e^{i\omega(t-t')} \\ \nonumber &\times& 
e^{-i[\delta_\alpha(t)-\delta_\alpha(t')]} 
 {\bf \Gamma}_\alpha {\bf G}^r(t,t') ,
\end{eqnarray}
where $\delta_\alpha(t)=\Delta_\alpha(t)-\Delta_g(t)$. 

The formula (\ref{general}) is manifestly {\sl gauge-invariant}: adding
an arbitrary time-dependent voltage to all (source, drain, and
gate) electrodes leaves  $I_\alpha(t)$ unaffected.
However, Eq.~(\ref{general}) {\sl per se}\ does not include 
{\sl displacement currents} \cite{footdis},
which are generally necessary for conservation of total charge 
in time-dependent quantum transport 
\cite{buttiker,fransson,polianski,polianski2}.  
 Moreover, they generate a driving-induced energy 
renormalization of the dot states \cite{buttiker,fransson}, which 
has to be determined self-consistently. 
On general grounds, Fransson \cite{fransson} has shown
that for arbitrary drive frequencies, the self-consistently 
evaluated displacement currents can be separated into independent 
(left and right) parts. Thus, they cannot contribute to the mixing 
amplitudes (\ref{gen}), and we shall disregard them from now on. 
Having established gauge invariance, we also put 
$\Delta_g(t)=0$ in what follows.

Using Eq.~(\ref{general}), it can now be shown that
in the absence of interactions,
frequency mixing is suppressed by the parametrically
small ratio ${\rm Tr}({\bf \Gamma}_{L/R})/D$, 
where $D$ is the electronic bandwidth in
the leads. Therefore, {\sl in the wide band limit for the leads, no 
frequency mixing occurs for a noninteracting quantum dot},  $J_{nm}=0$, 
regardless of how complicated its level structure may be.  
Technically, this follows because in the WBL the noninteracting GF
${\bf G}_0^r(t,t')$ is independent of the ac drive,
while ${\bf G}_0^<(t,t)$
depends on the left/right drive in an additive way only \cite{jauho}.
One therefore encounters only Fourier components $I_{nm}\ne 0$ 
for $n=0$ or $m=0$ in Eq.~(\ref{gen}).  The WBL for the leads is known to 
provide an excellent description when transport is dominated 
by states close to the Fermi level where ${\bf \Gamma}_\alpha$
has only weak energy dependence. 
Note that the same conclusion holds true for 
$[I_L(t)+I_R(t)]/2$, as the ``no mixing theorem'' applies to $I_L$ and $I_R$
separately. Below we focus on the $J_{nm}$ extracted from 
$I(t)=[I_L(t)-I_R(t)]/2$, and  adopt the WBL, where $J_{nm}\ne 0$ 
can only be caused by interactions.
Moreover, from now on we assume that only a single dot level is relevant
and take symmetric hybridization $\Gamma_L=\Gamma_R=\Gamma/2$.
The respective generalizations are straightforward but
do not yield new physics.  
The absence of mixing in the noninteracting limit may come as
a surprise, since even in the WBL the $IV$ curve of an undriven 
single-level dot is nonlinear. 
For instance, for a noninteracting resonant level at $T=0$, one finds
$I(V) = \frac{2e\Gamma}{h} \tan^{-1}(eV/2\Gamma)$.
While this nonlinearity allows to generate higher harmonics
 and rectification \cite{goldin1,polianski2}, 
it does not create finite mixing amplitudes in Eq.\ (\ref{gen}).
The latter arise from an effective ``cross-talk'' between the
source and drain electrodes, which
can only be mediated by interactions.  

The fact that interactions induce mixing can be demonstrated
on the simplest possible level by first-order perturbation theory in $U$
for the Anderson dot. The retarded
self energy is then given by $\Sigma^r(t,t') = U n(t) \delta(t-t')$, 
with the time-dependent dot occupation $n(t) = -iG_0^{<}(t,t)$. 
Expanding in Bessel functions, a closed expression for $n(t)$ in terms
of the free GF and Fermi functions follows. 
Notably, in the driven case, the dot occupation and hence $\Sigma^r$
become time-dependent, while for the time-independent case,
there is only a rigid shift of the dot level. 
The GF correction $\Delta G^r= G_0^r \Sigma^r G_0^r$ 
now generates mixing, and after some algebra, we 
obtain the mixing coefficients to first order in $U$
($n,m\neq 0$),
\begin{eqnarray} \label{full}
I_{nm} &=& iU \sum_{kl=-\infty}^\infty J_{k+n}(a_L) J_k(a_L)
J_{l+m}(a_R) J_{l}(a_R) \\   \nonumber
&\times& \nonumber \Bigl[
 F_a^{R}((l+m)\omega_R, l \omega_R) F_r^{L}
(-k\omega_L,-k\omega_L-m\omega_R) \\ \nonumber
&-& F_a^{L}((k+n)\omega_L, k \omega_L) F_r^{R}
(-l\omega_R,-l\omega_R-n\omega_L) \Bigr],
\end{eqnarray}
with the auxiliary functions
\[
F^{R/L}_{r/a} (\omega_1,\omega_2) = \Gamma \int \frac{d\epsilon}{2\pi}
f_{R/L}(\epsilon) G_0^r(\epsilon+\omega_1) G^{r/a}_0(\epsilon+\omega_2),
\]
which can be evaluated in closed (but lengthy) form.
They obey the symmetry relations
$F_r(\omega_1,\omega_2)=F_r(\omega_2,\omega_1)$ and 
$F_a(\omega_1,\omega_2)=F_a^*(\omega_2,\omega_1)$. 
Elementary inspection of Eq.~(\ref{full}) shows that for $V=0$,
$I_{nm} \to - I_{mn}$ for $\omega_R\to \omega_L$, 
as expected when exchanging source/drain contacts. 
Eq.\ (\ref{full}) nicely illustrates the basic mechanism: for $U\ne 0$, 
terms containing $\omega_L$ and $\omega_R$ appear in a multiplicative way, 
showing that mixing is due to ``cross-talk'' of the leads. 
For zero dc bias ($V=0$) and $\epsilon_0=0$, 
Eq.~(\ref{full}) and the above symmetry relations for $F_{r/a}$
imply $J_{nm}=0$
for all {\sl even} $n+m$.  Frequency mixing in this particle-hole
symmetric limit thus disappears, say,
 at the difference frequency 
$\omega_{1,-1}=\omega_L-\omega_R$, but is still present at $\omega_{2,-1}$. 
However, once $V\ne 0$ or $\epsilon_0\ne 0$, one always finds
mixing. To see this, it is instructive to consider
very small drive amplitudes $V^{ac}_{L,R}$ in Eq.~(\ref{full}). Upon
expanding the Bessel functions, the mixing amplitudes 
$J_{nm} \propto a_L a_R$ are non-zero only for $n,m=\pm 1$, and
the double sum \eqref{full} receives contributions only
from $k=0,-n$ and $l=0,-m$. Since $n+m$ is now always even,
there is no mixing unless $V\neq 0$ or $\epsilon_0\neq 0$.  
This simple calculation already demonstrates that mixing 
is generated by interactions. 

\begin{figure}
\includegraphics[width=0.45\textwidth]{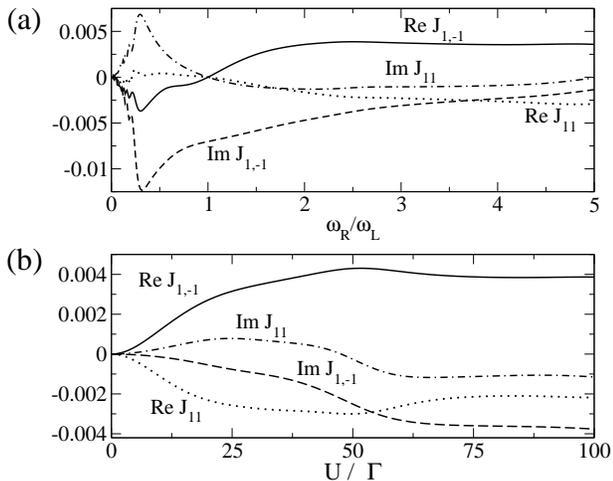}
\caption{ \label{fig1} 
Mixing amplitudes $J_{1,\pm 1}$ in the sequential tunneling regime,
in units of $2e\Gamma/\hbar$ for  $V=\epsilon_0=0, \omega_L=20\Gamma, 
V^{ac}_L=V^{ac}_R=50\Gamma$ and $T=5\Gamma$,  as a function
 of (a) $\omega_R/\omega_L$ for $U=100\Gamma$, and
(b) of $U/\Gamma$ for $\omega_R=50\Gamma$.
 }
\end{figure}

To study interaction effects beyond lowest order, we
next consider the {\sl sequential tunneling regime}
with $T>\Gamma$, where a master equation approach applies.
Here one evaluates the dynamics of the occupation probabilities $P_s(t)$
[with $0\leq P_s(t) \leq 1$ and $\sum_{s=1}^4 P_s(t)=1$]
for the four possible dot configurations $s$  
with energy $\epsilon_s$:
$s=1$ denotes the empty dot ($\epsilon_1=0$), $s=2,3$ the singly-occupied dot
 with spin up/down ($\epsilon_{2,3}=\epsilon_0$), and $s=4$ the
doubly occupied one ($\epsilon_4=2\epsilon_0+U$).  
Following standard steps \cite{bruder}, the master equation reads
 $\dot P(t) = \sum_{s',\alpha=L/R} \left[ K^\alpha_{ss'}(t) P_{s'}(t) -
K^\alpha_{s's}(t) P_s(t) \right]$
with transition rates ($s\ne s'$, where
 $K^\alpha_{ss}=-\sum_{s'\ne s} K^\alpha_{ss'}$)
\begin{eqnarray}\label{rates}
K_{ss'}^\alpha(t)&=& \Gamma_\alpha {\rm Re}  \sum_{k,q=-\infty}^\infty
 e^{iq\omega_\alpha t} J_k(a_\alpha) J_{k+q}(a_\alpha) \\ 
&\times& \nonumber \sum_\pm  N_\pm g_\pm 
(\pm[\epsilon_s-\epsilon_{s'}]+k\omega_\alpha-\alpha eV/2),
\end{eqnarray}
where $N_{+}=\sum_{\sigma=\uparrow,\downarrow} |\langle 
s|d_\sigma^{\dagger}| s'\rangle |^2$, 
$N_-$ follows by replacing the dot fermion operator
$d^\dagger_\sigma\to d_\sigma$, and
\begin{equation}
g_\pm(\epsilon)=f(\epsilon) 
\pm \frac{i}{\pi} \ln\frac{D}{2\pi T} \mp \frac{i}{\pi} {\rm Re}
\ \psi\left(\frac12 + \frac{i\epsilon}{2\pi T} \right),
\end{equation}
where $\psi$ is the digamma function, implying logarithmic
divergencies \cite{costi}.
After Fourier expansion as in Eq.~(\ref{gen}),
the master equation leads to an algebraic equation.
Numerical solution obtains the Fourier coefficients $P_{s,nm}$
and hence the $J_{nm}$ from 
$I_\alpha(t)=\sum_{s\ne s'} \theta_{ss'}K_{ss'}^\alpha(t) P_{s'}(t)$, 
where $\theta_{ss'}=\pm 1$ for $s \gtrless s'$.
These equations allow to reproduce known results for the 
sequential tunneling current under a dc bias 
and for the ac driven case with $\omega_R=\omega_L$ \cite{bruder}.  
For $U=0$, in accordance with our discussion above,
no frequency mixing is found from the master equation,
while $J_{nm}\ne 0$ for $U\ne 0$. 
To ensure consistency, we have checked that for small $U$,
the master equation reproduces the perturbative result 
(\ref{full}) taken at high $T$ or large $V$.

Figure \ref{fig1} shows the mixing amplitudes $J_{1,\pm 1}$ 
for $V=\epsilon_0=0$ as a function of $U/\Gamma$ and $\omega_R/\omega_L$.
Note that for $\omega_R=\omega_L$, we correctly find
$J_{11}=0$ as enforced by the $V=0$ symmetry $I_{nm}=-I_{mn}$ under 
exchange of $\omega_R$ and $\omega_L$.  At this point, ${\rm Re} J_{1,-1}$
also vanishes, while ${\rm Im} J_{1,-1}\ne 0$ does not  generate
current, $I(t)\propto \sin[(\omega_R-\omega_L)t]=0$.
Remarkably, the mixing amplitudes display characteristic 
features (peaks or steps) at certain ratios $\omega_R/\omega_L$,
cp.~for $\omega_R/\omega_L=1/3$ in Fig.~\ref{fig1}(a). 
Such features can be rationalized in simple terms
by noting that the ac voltages correspond to 
photon-assisted side peaks in the 
dot's density of states, located at energies $\epsilon_0\pm \omega_{nm}$
and  $2\epsilon_0+U\pm \omega_{nm}$ with arbitrary integers $n,m$.
Once one of those energies hits the Fermi level  
(which is located at $\epsilon_0=0$ in Fig.~\ref{fig1}),
transition amplitudes are resonantly enhanced, and 
mixing becomes particularly efficient. To give an example, such
resonances occur for $\omega_R/\omega_L = (U/\omega_L-n)/m$, and
 in Fig.~\ref{fig1}(a), where $U/\omega_L=5$, the feature
at $\omega_R/\omega_L=1/3$ corresponds to $(n,m)=(4,3)$.
Which of these commensurability features
(indexed by $n,m$) will actually show up in the mixing current is
primarily determined by the drive amplitudes $V^{ac}_{L/R}$. 
For the rather large $V^{ac}_{L/R}$ taken in Fig.~\ref{fig1},
large $(n,m)$ have to be taken into account, resulting in the rather
complicated dependence of the mixing amplitudes
on $\omega_R/\omega_L$ observed for $\omega_R/\omega_L<1/3$.
The data in Fig.~\ref{fig1}(a) show that mixing disappears
in the limit $\omega_L\to \infty$. We also observe
that mixing disappears in the opposite limit $\omega_R/\omega_L
\to\infty$, where one can effectively average over the fast
$\omega_R$ oscillations and ends up with a monochromatic situation again.
Finally, additional calculations (not shown) reveal 
that like in the small-$U$ case, there is no mixing ($J_{nm}=0$)
at the particle-hole symmetric point,
$\epsilon_0=-U/2$ and $V=0$, for all even $n+m$.
Before turning to the case of e-ph interactions, we
briefly comment on the {\sl Kondo regime}, 
realized at low temperatures for $U\gg \Gamma$ and $\epsilon_0\approx -U/2$,
where one can map the Anderson dot to a spin-$1/2$ impurity problem.
At the special Toulouse point, an exact solution of the mixing problem 
can then be obtained from Ref.~\cite{schiller}. This solution
will be discussed elsewhere, but
by taking $T=V=0$, mixing is seen to disappear, 
in accordance with the Fermi liquid nature of the Kondo fixed point.
At finite $T,V$, however, we again find nonzero mixing amplitudes.

\begin{figure}
\includegraphics[width=0.45\textwidth]{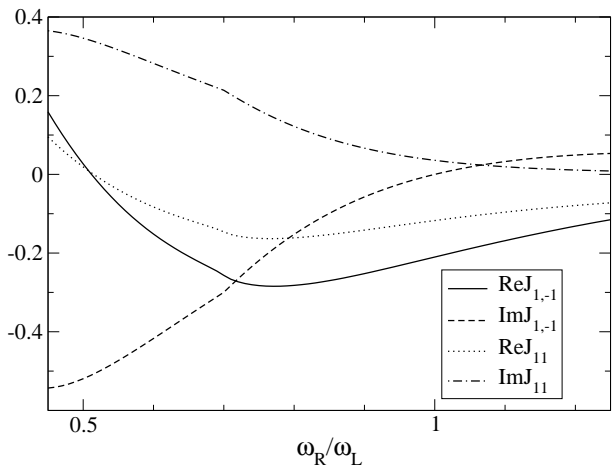}
\caption{ \label{fig2} Mixing amplitudes $J_{nm}$
for phonon-mediated interactions and small drive amplitudes $V^{ac}_{L/R}$,
see Eq.~(\ref{phonon}) expanded to order $V^{ac}_L V^{ac}_R$.  
Parameters:  $T=\epsilon_0=0, \Omega=\Gamma, V=0.8\Gamma, 
\omega_L=2\Gamma.$ The $J_{nm}$
are given in units of $\lambda^2 V_L^{ac} V_R^{ac}/(2\pi \Gamma)^3$. 
 }
\end{figure}

Next we discuss a spinless level (fermion operator $d$),
with e-ph coupling $\lambda$ 
to a phonon mode $Q=b+b^\dagger$ of frequency $\Omega$
(boson operator $b$).
The dot is described by
$H_{\rm dot}= [\epsilon_0+\lambda Q] d^\dagger d + \Omega b^\dagger b$,
and connected to leads as for the Anderson dot. We have computed the 
mixing amplitudes $J_{nm}=I_{nm}+I^*_{-n,-m}$
by second-order perturbation theory in $\lambda$, 
with the result $I_{nm}= [I^{+}_{nm} - I^{-}_{mn}]/2$, where
\begin{eqnarray}  \nonumber 
I^{\alpha}_{nm} &=& i(\lambda \Gamma/2\pi)^2 \sum_{k,l}
J_{l+n}(a_\alpha) J_{l}(a_\alpha) J_{k+m}(a_{-\alpha}) J_k(a_{-\alpha}) 
\\ \nonumber &\times& \int d\omega \int d\epsilon
f_\alpha(\omega) f_{-\alpha}(\epsilon)
 D^r_0(\omega-\epsilon-l\omega_\alpha -k\omega_{-\alpha}) \\
\nonumber &\times&
G_0^r (\epsilon+(k+m)\omega_{-\alpha}) G_0^a(\epsilon+k\omega_{-\alpha})
G_0^r(\omega-l\omega_\alpha) 
 \\  &\times& G_0^r(\omega-l\omega_\alpha+m\omega_{-\alpha}) ,
\label{phonon}
\end{eqnarray}
where $D_0^r(\omega) = 2\omega/[(\omega+i0^+)^2-\Omega^2]$.
{}From these expressions, one shows that for $V=\epsilon_0=0$, again
$J_{nm}\ne 0$ only for even $n+m$. 
For $\Omega \gg \Gamma$, the high phonon frequency allows
to approximate the boson propagator by a constant, and one essentially
comes back to the Anderson dot result (\ref{full}) where mixing 
was established above.  In Fig.~\ref{fig2}, we show the mixing
amplitudes for $\Omega=\Gamma$ and small drive amplitudes in Eq.~(\ref{phonon}).

Finally, we comment on how frequency mixing 
could be verified experimentally, and on consequences 
 for applications. Working quantum-dot frequency mixers 
 \cite{Cleland,Buehler} operate in the adiabatic regime at  rather low 
$\omega_{R/L}$ (no photon-assisted effects), probably outside the 
applicability range of our theory \cite{footf}. 
The measured dot capacitance $C=550$ aF \cite{Cleland} 
puts $U=e^2/2C$ around $35$ GHz, while 
$\Gamma = 2\pi/R C\approx 13$ GHz follows from the 
measured resistance $R=850$~k$\Omega$, implying $U/\Gamma\approx 2.7$. 
By comparing to Fig.~\ref{fig1}, a large mixing signal 
is expected for $U/\Gamma\approx 50$, 
which may be achieved for the same $C$ by increasing $R$
to $R\approx 23$~M$\Omega$, leading to $\Gamma=500$~MHz. 
Choosing $\omega_{R/L}$ in the GHz range, see Fig.~\ref{fig1},
should then put the device into
an experimentally accessible regime, where 
our theory applies and predicts strong frequency mixing.
In fact, one may be able to generate THz waves by such a mixer.
In the present work, we have shown that mixing
is possible for $V=0$, where the dc current vanishes and hence
a very small noise level is expected. 
Existing experiments \cite{Cleland} involved finite dc current,
causing large noise levels and dissipation effects in the detection electronics.
Both these problems can be considerably reduced by using a correlated dot
near $V=0$, where interactions provide the required nonlinearity.

To conclude, the theory of harmonic frequency mixing in interacting
quantum dots has been given. 
In the wide-band limit for the leads, a ``no mixing 
theorem'' can be established, stating that mixing requires the
presence of interactions.  For both e-e and e-ph
interactions, we have then shown that mixing is indeed generated,
and provided detailed quantitative predictions for the mixing amplitudes. 
We hope that these findings stimulate experiments and further
theoretical developments.

We thank M.~Devoret for inspiring this work, and K. Flensberg, 
A.P. Jauho, Yu. Nazarov, J. Paaske, M. Polianski, and 
H. Schoeller for discussions.
This work was supported by the DFG SFB TR 12 and by the
ESF network INSTANS. A.O.G. thanks the Humboldt foundation
for a Friedrich-Wilhelm-Bessel award enabling his extended stay
in D\"usseldorf.


\begin{thebibliography}{99}
\bibitem{footf}
Our approach is valid for frequencies $\omega_{L/R}$ smaller than the
plasma frequency in the leads (typically a few THz),
but larger than $\Gamma$ (typically a few hundred MHz).
\bibitem{footi}
In principle, the parameters of the ``noninteracting'' 
model have to be computed from a self-consistent mean-field
treatment of the interactions.  We here discuss
interaction phenomena beyond such single-particle effects.
\end{thebibliography}
\end{document}